\documentclass[letterpaper,12pt]{article}   %% LaTeX 2e (preferred)
\usepackage{verbatim}

\usepackage[]{graphicx}
\usepackage{amsmath}
\usepackage{amssymb}
\usepackage{epsfig}
\usepackage{color}
\usepackage{subfigure}

 \advance \textwidth by 1cm
\def\blue{\textcolor{blue}}

\def\##1{{\bf #1}}
\def\=#1{\underline{\underline #1}}

\def\.{\mbox{ \tiny{$^\bullet$} }}

\def\r#1{(\ref{#1})}

\def\eps{\varepsilon}

\def\epso{\eps_{0}}
\def\lambdao{\lambda_{ 0}}
\def\muo{\mu_{ 0}}
\def\ko{k_{ 0}}

\def\epsd{\eps_d}
\def\epssil{\eps_{s}}

\def\pinc{{\mathbf p}^+}
\def\pref{{\mathbf p}^-}

\def\ux{\hat{\#u}_x}
\def\uy{\hat{\#u}_y}
\def\uz{\hat{\#u}_z}

\begin{document}

\begin{center}

\textbf{Grating-coupled  excitation of the Uller--Zenneck surface wave in the optical regime}\\

 {\it Muhammad Faryad} and
{\it Akhlesh Lakhtakia}\\
{Nanoengineered Metamaterials Group (NanoMM), Department of Engineering Science and
Mechanics, 
Pennsylvania State University, University Park, PA  16802-6812, USA}

\end{center}

\begin{abstract}
The excitation of the Uller--Zenneck surface wave in the optical regime was theoretically investigated for planar as well as periodically corrugated interfaces of two homogeneous, isotropic dielectric materials, with only one of the two being dissipative. A practical configuration involving the planewave illumination of a planar interface
of the two partnering materials was found to be unsuitable for experimental confirmation
of the existence of this surface wave. But, when the interface was  periodically corrugated, the Uller--Zenneck wave was found to be excited over a wide range of the angle of incidence. Air and crystalline silicon were identified as suitable partnering materials for experiments in the visible and ultraviolet spectral regimes.
\end{abstract}

\section{Introduction}
In 1907, Zenneck published a theoretical analysis of  a radio-frequency   surface wave guided by the planar interface of air and ground, both assumed homogeneous and isotropic  \cite{Zenneck}. Subsequently, an electromagnetic surface wave guided by the interface of two homogeneous and isotropic dielectric materials of which only one is dissipative came to be called the Zenneck wave. A 1903 analysis of the same type of wave guided by the interface of a nondissipative dielectric material and (dissipative) seawater by Uller {\cite{Uller}} appears to have become obscure, except for  citations by  both Zenneck  \cite{Zenneck} and Collin~\cite{Collin_book}. Since Uller not only obtained but also solved the dispersion equation of the surface wave, albeit under the special conditions of seawater being significantly conductive
and the (real) relative permittivity of seawater being much larger than that of its partnering material (air?), we think it is appropriate to name this surface wave after both Uller and Zenneck.

Sommerfeld~\cite{Sommerfeld1909,Sommerfeld1926} provided a rigorous mathematical analysis of   the Uller--Zenneck wave; see also a review by Wait \cite{Wait1998}. As the ground is not metallic but is a dissipative dielectric material, the  Uller--Zenneck wave must be distinguished from the surface plasmon-polariton (SPP) wave that is usually taken to be  guided by the interface of a lossless dielectric material and a metal, both assumed homogeneous and isotropic~\cite{Maier07,Homola_book}. 
 
 Although sometimes ignored \cite{Jeon,Navarro-Cia}, the distinction between SPP and Uller--Zenneck waves has a significant consequence. 
Unlike an SPP wave, whose phase speed is smaller than that of a plane wave in the partnering dielectric material, the phase speed of an Uller--Zenneck wave is usually larger than the phase speeds of  plane waves in both  partnering dielectric materials. Accordingly, unlike the SPP wave \cite{Homola_book}, the Uller--Zenneck wave can be excited without  the use of a coupling prism or a surface-relief grating. All that is needed is to illuminate the guiding planar interface by a highly obliquely incident plane wave. 

But, that very same characteristic leads to difficulties in distinguishing the Uller--Zenneck wave from  other waves in the radio-frequency regime \cite{Wait1998}.
Experimental observation of  the Uller--Zenneck wave in the same spectral regime~\cite{Baibakov} remains mired in
confusion~\cite{Datsko}. Part of the reason for confusion about the Uller--Zenneck wave is an infelicitous connection with the   Brewster phenomenon~\cite{Wait_book,Collin_book}.

\begin{figure}[!ht]
\begin{center}
\includegraphics[width=4.5in]{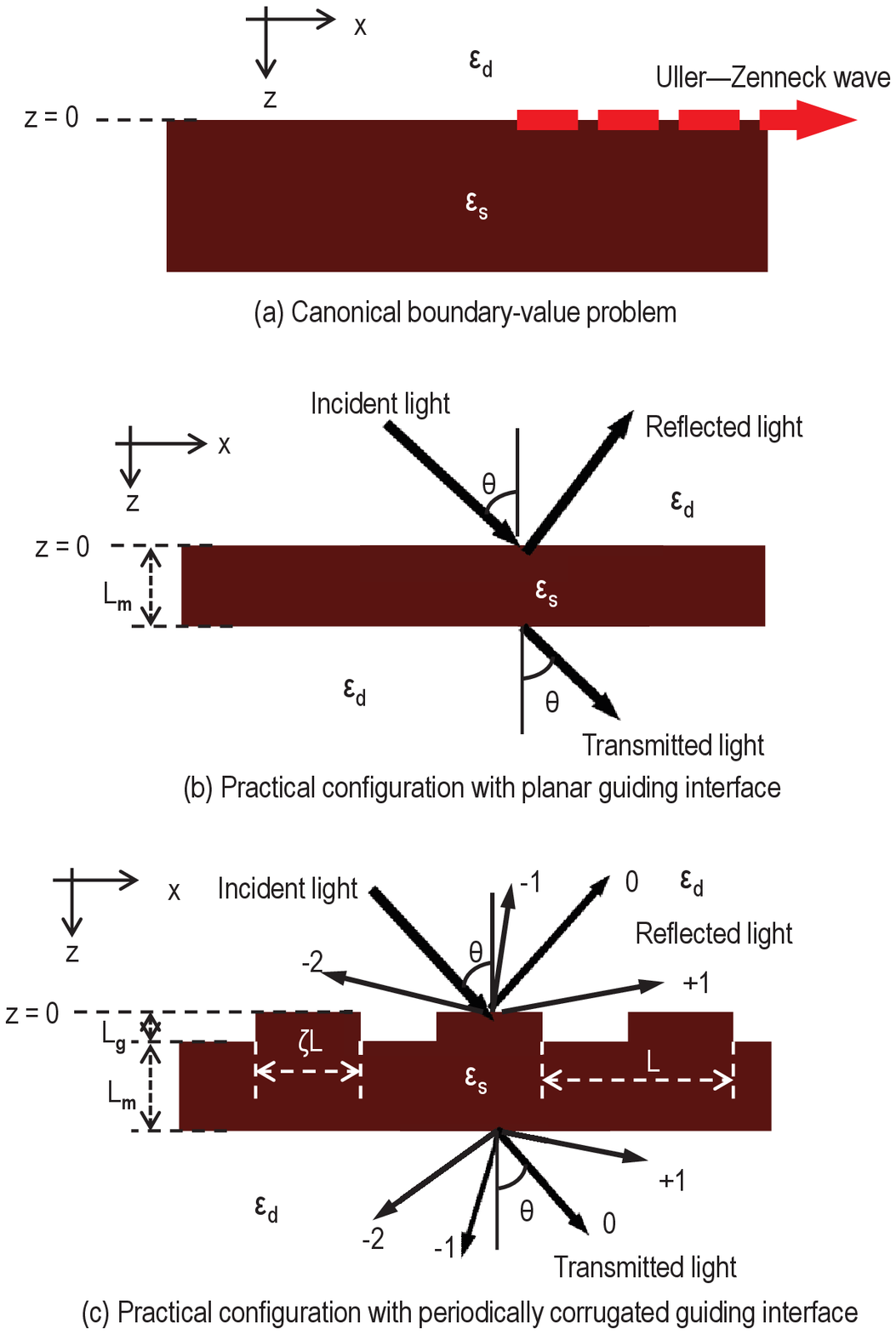}
\caption{(a) Schematic of the canonical boundary-value problem. The Uller--Zenneck wave is guided by the interface $z=0$. (b) Excitation of the Uller--Zenneck wave by a plane wave that is obliquely incident at the interface $z=0$, when the dissipative partnering material has a large but finite thickness. (c)  
{Schematic of the grating-coupled configuration. Same as (b), but the guiding interface is periodically corrugated.} The specular components of the reflected and transmitted fields
are labeled as $0$, whereas their non-specular components are labeled by non-zero integers.
 }
\label{schematics}
\end{center}
\end{figure}

Theory, on the other hand, is unequivocal.  Uller \cite{Uller}, Zenneck \cite{Zenneck}, and Sommerfeld \cite{Sommerfeld1909,Sommerfeld1926} had investigated the
 canonical boundary-value problem depicted schematically in Fig.~\ref{schematics}(a). In this figure,  the   nondissipative dielectric partnering material with relative permittivity $\epsd$ (${\rm Re}(\epsd)>0$,
 ${\rm Im}(\epsd)=0$) occupies the half space $z<0$; the    dissipative dielectric partnering material with relative permittivity $\epssil$ (${\rm Re}(\epssil)>0$,
 ${\rm Im}(\epssil)>0$) occupies the half space $z>0$; and the Uller--Zenneck wave propagates parallel to the $x$ axis, decays as $\vert{z}\vert\to\infty$, and does not depend on $y$. The wave has to be 
{$p$ polarized} (i.e., $\uy\.\#E=0$ and $\ux\.\#H=\uz\.\#H=0$, where the Cartesian unit vectors are identified as $\ux$, $\uy$, and $\uz$). The dispersion equation  is readily solvable~\cite{PMLbook}.

However, the canonical boundary-value problem is not implementable practically. At the very least, the dissipative partnering material has to have a large but finite thickness $L_m$, as shown schematically in Fig.~\ref{schematics}(b), and a $p$-polarized  plane wave has to be incident on the interface $z=0$ from
 the half space $z<0$. Instead of a $p$-polarized plane wave, an angular spectrum of $p$-polarized plane waves emanating from a finite source can be used \cite{HillWait,Hill1980}. Air has been taken as the nondissipative partnering material in most radio-frequency experiments \cite{Baibakov}. But the evidence for the excitation of the Uller--Zenneck wave in this configuration is ambiguous \cite{Datsko,Wait1998}.
 
 The grating-coupled configuration \cite{Faryad2} provides an alternative. As shown in 
 Fig.~\ref{schematics}(c), the  interface of the two partnering materials is  periodically corrugated
 in this configuration.
Although some evidence is available from a finite-element simulation  \cite{Hendry} in the radio-frequency regime, with the periodic corrugations modeled as an impedance plane with periodically varying surface impedance,  no comparison was made in that study against the underlying canonical boundary-value problem and the localization of the surface wave was not explicitly  demonstrated. 
 
 Since the grating-coupled configuration is practicable   in the optical regime \cite{Homola_book,PFLHL}, we decided to examine theoretically if unambiguous experimental evidence of the Uller--Zenneck wave can be obtained therefrom. Therefore, the grating-coupled configuration  is the focus of this paper, although the underlying canonical boundary-value problem and  excitation using the planar interface were also  investigated.   

The plan of this paper is as follows: The canonical boundary-value problem is briefly  discussed and numerical results are presented  in Sec.~\ref{canonical}.  Excitation of the Uller--Zenneck wave by a plane wave obliquely incident at the interface $z=0$, when the dissipative partnering material has a large but finite thickness,   is discussed in Sec.~\ref{tkrconfig}. The numerical results for the grating-coupled configuration are  discussed in Sec.~\ref{gratingconfig}. Concluding remarks are presented in Sec.~\ref{conc}.
An $\exp(-i\omega t)$  dependence on time $t$ is implicit, with $\omega$ denoting the angular frequency and $i=\sqrt{-1}$. Vectors are in boldface.  The free-space wavenumber and  wavelength  are denoted by $\ko=\omega\sqrt{\epso\muo}$ and $\lambdao=2\pi/\ko$, respectively, with $\muo$ and $\epso$ being  the permeability and permittivity of free space{, respectively.}

\section{Canonical boundary-value problem}\label{canonical}
Let us begin with the  canonical boundary-value problem, shown schematically in Fig.~\ref{schematics}(a). The surface wave has to be $p$ polarized. If it is taken to propagate parallel to the $x$ axis in the $xz$ plane, 
 the electric field phasors in the two contiguous  half spaces may be written as 
\begin{equation}
\#E(\#r)= a_{p}\left( \frac{\alpha_{d}\ux+q\uz}{\ko n_d}
\right)
\exp\left[i\left(qx-\alpha_{d}z\right)\right]\,,\quad z < 0\,,
\label{Eminus}
\end{equation}
and
\begin{equation}
\#E(\#r)= b_{p}\left(  \frac{-\alpha_{s}\ux+q\uz}{\ko n_s}
\right)
\exp\left[i\left(qx+\alpha_{s}z\right)\right]\,,\quad z > 0\,,
\label{Eplus}
\end{equation}
where $n_d=\sqrt{\epsd}>0$, $n_s=\sqrt{\epssil}$,
$q^2+\alpha_{s}^2=\ko^2\, \epssil$, and $q^2+\alpha_{d}^2=\ko^2\epsd$. Furthermore, 
$q$ is complex valued, ${\rm Im}(\alpha_{d})>0$ for attenuation as $z\to-\infty$, and  ${\rm Im}(\alpha_{s})>0$ for attenuation as $z\to\infty$. Finally $a_{p}$ and $b_p$ are unknown scalars with the same units as the electric field. 

The dispersion equation of the Uller--Zenneck wave can be found after enforcing the standard boundary conditions at the interface $z=0$. The solution of the dispersion equation is
\begin{equation}
\label{dispsol}
q=\ko\sqrt{\frac{\eps_d\eps_{s}}{\eps_d+\eps_{s}}}\,,
\end{equation}
where
\begin{equation}
\alpha_d=\ko\frac{\eps_d}{\sqrt{\eps_d+\eps_s}}\,,\quad
\alpha_{s}=\ko\frac{\eps_s}{\sqrt{\eps_d+\eps_s}}\,.
\end{equation}

\begin{figure}[!ht]
\begin{center}
\includegraphics[width=3.5in]{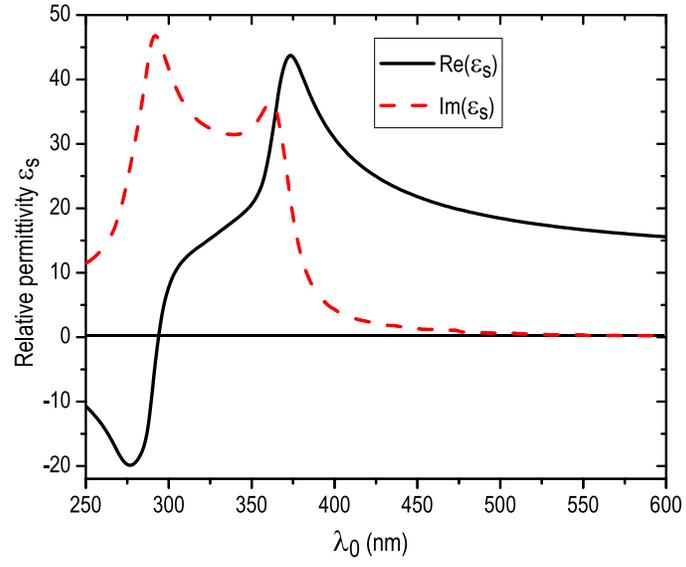}
\caption{Real and imaginary parts of the relative permittivity $\epssil$ of crystalline silicon
as functions of $\lambdao$ \cite{Palik_book}. }
\label{silicondata}
\end{center}
\end{figure}

For illustrative numerical results throughout this paper, we identified crystalline silicon as the  dissipative partnering material and air as the nondissipative partnering  material. The relative permittivity   $\epssil$ of crystalline silicon is presented
in Fig.~\ref{silicondata}
as a function of  $\lambdao$ \cite{Palik_book}, whereas  $\epsd$ was taken to be unity.

\begin{figure}[!ht]
\begin{center}
\includegraphics[width=3.5in]{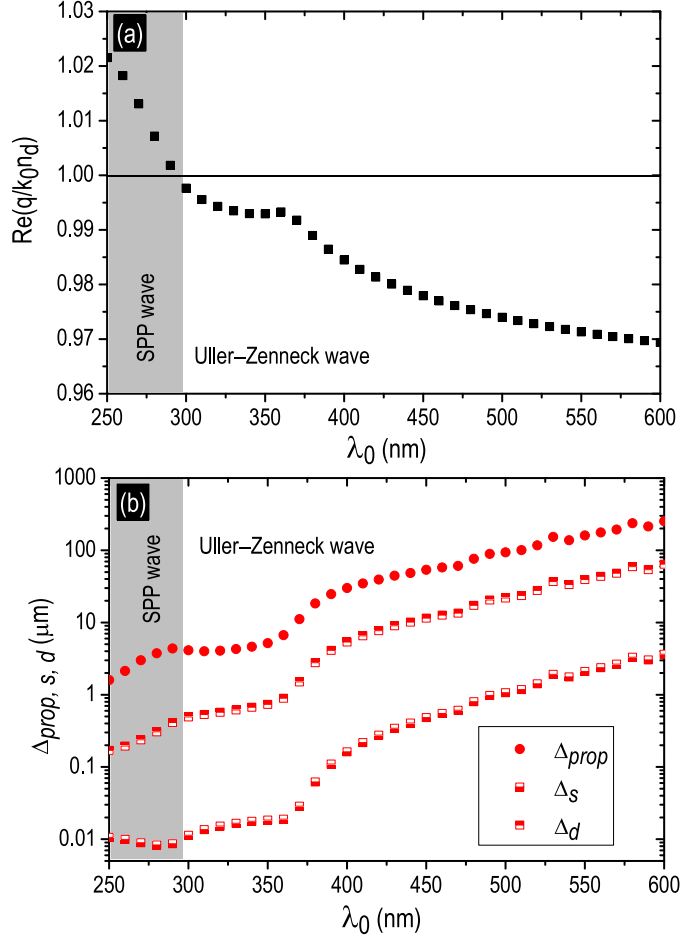}
\caption{Canonical boundary-value {problem.} (a) Spectrum of ${\rm Re}(q)/\ko{n_d}$ {for} the surface wave guided by the planar interface of air and crystalline silicon. (b) Spectrums
of $\Delta_{prop}$, $\Delta_d$, and $\Delta_s$. The surface wave is  an SPP wave for $\lambdao<293$~nm and an Uller--Zenneck wave for $\lambdao\geq294$~nm.}
\label{can}
\end{center}
\end{figure}

Figure~\ref{can}(a) presents the spectrum of ${\rm Re}(q)/\ko{n_d}$ for $\lambdao\in[250,600]$~nm.  
As silicon is  metallic for $\lambdao\leq293$~nm because ${\rm Re}(\epssil)<0$,  the surface wave must be classified as an SPP wave. However, silicon is  a dielectric material, though very dissipative, for
$\lambdao\geq294$~nm because ${\rm Re}(\epssil)>0$; hence, the surface wave must then be classified as an Uller--Zenneck wave.
Let us also note that  ${\rm Re}(q)>\ko{n_d}$ for SPP waves and  ${\rm Re}(q)\lesssim\ko{n_d}$ for Uller--Zenneck waves.

The propagation length $\Delta_{prop}=1/{\rm Im}\left(q\right)$   is presented in Fig.~\ref{can}(b)
as a function of $\lambdao$. Whereas $\Delta_{prop}<5$~$\mu$m for the SPP wave,
 the propagation length of the Uller--Zenneck wave lies between
$5$ and $700$~$\mu$m. In general, $\Delta_{prop}$ increases as $\lambdao$ increases since Im$(\epssil)$ decreases. The increase of the propagation length is also accompanied by  increases in the penetration depth $\Delta_d =1/{\rm Im}\left(\alpha_{d}\right)$ of the surface wave in the
nondissipative partnering material
 and the penetration depth
$\Delta_s= 1/{\rm Im}\left(\alpha_{s}\right)$ into the 
dissipative partnering material, as also shown in Fig.~\ref{can}(b).

Although the canonical problem is not practically implementable, predictions of ${\rm Re}(q)$ provided
by its solution are valuable in designing practical configurations. With air as the nondissipative partnering material, two practical configurations are considered next. In both configurations, the dissipative partnering material is present as a slab of finite thickness. A  $p$-polarized plane wave is obliquely incident
on one face of this slab. The face can be either planar (Sec.~\ref{tkrconfig}) or periodically corrugated
(Sec.~\ref{gratingconfig}).

\section{Practical configuration with planar guiding interface}\label{tkrconfig}
As shown in Fig.~\ref{schematics}(b), let the half spaces $z<0$ and $z>L_m$ be filled with a homogeneous
material of relative permittivity $\epsd$, while the region $0<z<L_m$ is occupied by a
homogeneous material of relative permittivity $\epssil$. A  $p$-polarized plane wave is obliquely incident
on the plane $z=0$, its wave vector making an angle $\theta$ with respect to the $z$ axis. Equation~\r{dispsol} predicts that a surface wave will be excited when $\theta=\theta^{\rm C}
=\sin^{-1}\left[{\rm Re} (q)/\ko{n_d} \right]$. 
 
The electric field phasor in the half space $z < 0$ is given by
\begin{eqnarray}
\nonumber
\#E(\#r)&=& \left(-\ux\cos\theta+\uz\sin\theta\right)
\exp\left[i\ko{n_d}\left( x\sin\theta+ z\cos\theta\right)\right] \\
&&
+r_{p}\left(\ux\cos\theta+\uz\sin\theta\right)
\exp\left[i\ko{n_d}\left(x\sin\theta -z\cos\theta \right)\right]\,, {\quad}z<0\,,
\end{eqnarray}
where $r_p$ is the reflection coefficient. In the half space $z>L_m$, the electric field
phasor is given by
\begin{equation}
\#E(\#r)= t_{p} \left(-\ux\cos\theta+\uz\sin\theta\right)
\exp\left\{i\ko{n_d}\left[x\sin\theta+\left(z-L_m\right)\cos\theta\right]\right\}\,, {\quad} z> L_m,
\end{equation}
where $t_p$ is    the  transmission
coefficient.  The reflectance $R_{p}=\left|  r_{p}  \right|^2$, 
the transmittance $T_{p}=\left|  t_{p}  \right|^2$, and the absorptance
$A_p=1-\left(R_{p}+T_{p}\right)$
can be computed as functions of $\lambdao$ and $\theta$ for any value of $L_m$ using a textbook procedure
\cite{BW}.

\begin{figure}[!ht]
\begin{center}
\includegraphics[width=5.5in]{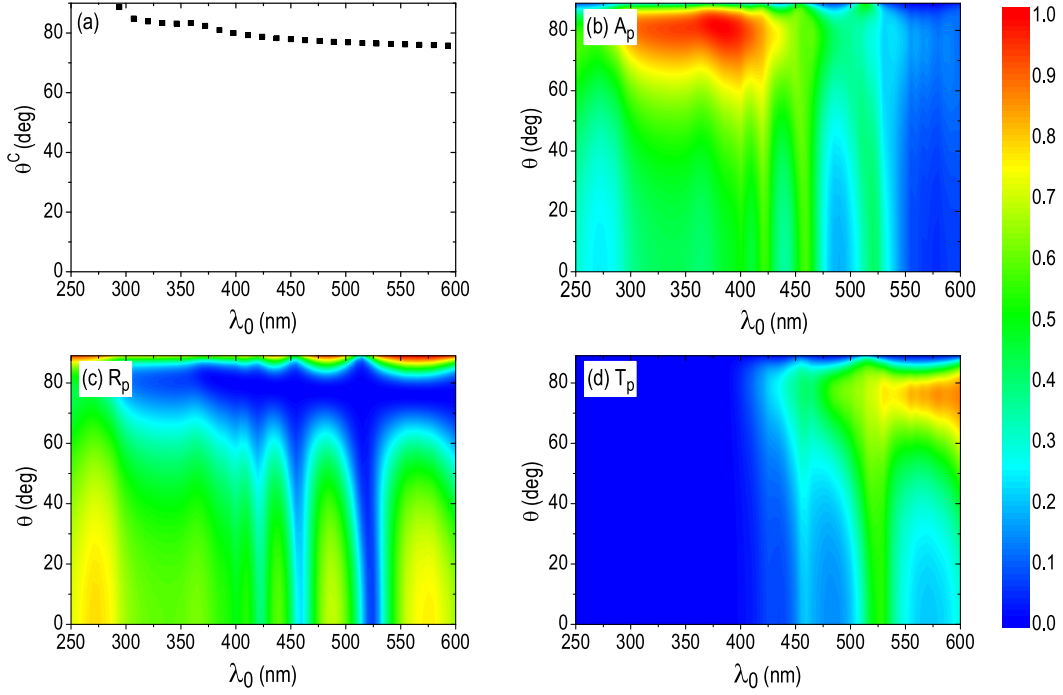}
\caption{Practical configuration  with planar guiding {interface.} (a)   Spectrum of $\theta^{\rm C}=\sin^{-1}\left[{\rm Re} (q)/\ko{n_d} \right]$ when the excitation of the Uller--Zenneck wave is
predicted by the solution of the canonical boundary-value problem.  (b) Absorptance $A_p$, 
(c) reflectance $R_{p}$, and (d) transmittance $T_{p}$ as  functions of $\lambdao$ and $\theta$  when $L_m=250$~nm.
}
\label{tkr}
\end{center}
\end{figure}

The plot of $\theta^{\rm C}$
as a function of $\lambdao$ for air and crystalline silicon as the partnering materials
is presented in Fig.~\ref{tkr}(a). For $L_m=250$~nm,  $A_p$, 
$R_{p}$, and   $T_{p}$ as  functions of $\lambdao$ and $\theta$  are presented
in Figs.~\ref{tkr}(b), (c), and (d), respectively. No sharp absorptance band---that could signify
the excitation of  a surface wave \cite{Hall2013}---is present
in Fig.~\ref{tkr}(b). Neither are any similar signatures of the excitation of surfaces waves present in 
Figs.~\ref{tkr}(c) and  (d). The same conclusions were drawn for other values of
$L_m\in\left[100,1000\right]$~nm. 

Parenthetically, the wide absorptance band in Fig.~\ref{tkr}(b) delineated approximately by $\lambdao\in\left[275,425\right]$~nm and $\theta\in\left[60^\circ,85^\circ\right]$ does not indicate surface-wave excitation. For any fixed value of $\lambdao$, the center  of the $\theta$-range of this band is significantly different
from $\theta^{\rm C}$. Moreover, the $\theta$-range is far too broad to signify the excitation of
any surface wave.

This practical configuration is the one employed for most experimental investigations of
the Uller--Zenneck wave. Figure~\ref{tkr} leads to the conclusion that this configuration is not appropriate
to  unambiguously confirm the existence of the Uller--Zenneck wave.

\section{Practical configuration with periodically corrugated  guiding interface}\label{gratingconfig}
Let us now consider the excitation of the Uller--Zenneck wave in the grating-coupled configuration, shown schematically in Fig.~\ref{schematics}(c). The regions $z<0$ and $z>L_t=L_g+L_m$ are occupied by a homogeneous material of relative permittivity $\epsd$, the region  $L_g< z< L_t$ is occupied by a homogeneous material of relative permittivity $\epssil$, and the region $0< z<L_g$ contains a rectangular grating  of period $L$ along the $x$ axis and  duty cycle  $\zeta \in(0,1)$. Let a $p$-polarized plane wave be obliquely incident upon the grating. The wave vector of the incident plane wave lies wholly in the $xz$ plane and is oriented at an angle $\theta$  with respect to the $z$ axis. 

The electric field phasor in the half space $z<0$ is adequately represented by \cite{Faryad2,PMLbook}
\begin{eqnarray}
\nonumber
\#E(\#r)&=& \left(-\ux\cos\theta+\uz\sin\theta\right)
\exp\left[i\ko{n_d}\left( x\sin\theta+ z\cos\theta\right)\right]\\
&&+
\sum_{n=-N_t}^{N_t} r_{p}^{(n)}\pref_n
\exp\left[i\left(\kappa^{(n)}x-\alpha^{(n)}z\right)\right]
\,,{\quad}  z\leq0\,.\label{elower}
\end{eqnarray}
Here, $N_t>0$ is a sufficiently large integer, $r_p^{(n)}$ is the amplitude of  the Floquet harmonic of 
 order $n\in \left\{0,\pm1,\pm2,...,\pm N_t \right\}$ in the reflected field, with
 \begin{eqnarray}
\kappa^{(n)}&=&\ko{n_d}\sin\theta+2{\pi}n/L\,,
\\
\alpha^{(n)}&=&
\begin{cases}
+\sqrt{\ko^2\epsd-{(\kappa^{(n)})}^2}\,,& \ko^2\epsd\geq {(\kappa^{(n)})}^2\\[5pt]
+i\sqrt{{(\kappa^{(n)})}^2-\ko^2\epsd}\,,& \ko^2\epsd< {(\kappa^{(n)})}^2
\end{cases}\,,
\\
{\#p}_n^\pm&=&\frac{\mp\alpha^{(n)}\ux+\kappa^{(n)}\uz}{\ko {n_d}}
\,.
\end{eqnarray}
In the half space $z>L_t$, the electric field
phasor is given by
\begin{equation}
\#E(\#r)=\sum_{n=-N_t}^{N_t}  t_{p}^{(n)} \pinc_n
\exp\left\{i\left[\kappa^{(n)}x+\alpha^{(n)}\left(z-L_t\right)\right]\right\}\,, {\quad}z> L_t\,, 
\end{equation}
where $t_p^{(n)}$ is the amplitude of  the Floquet harmonic of 
 order $n$ in the transmitted field.  Whereas $n=0$ 
 identifies the specular components of the reflected and transmitted fields, the non-specular components are identified by $n\ne0$. 
 
 The reflectances and transmittances of order
$n$ are defined as
\begin{equation}
R_{p}^{(n)}=\vert{r_p^{(n)}}\vert^2\, \frac{{\rm Re}\left(\alpha^{(n)}\right)}{\alpha^{(0)}}\,,
\quad
T_{p}^{(n)}=\vert{t_p^{(n)}}\vert^2 \,\frac{{\rm Re}\left(\alpha^{(n)}\right)}{\alpha^{(0)}}\,,
\end{equation}
respectively;
the total reflectance and the total transmittance as
\begin{equation}
R_p=\sum_{n=-N_t}^{N_t} R_{p}^{(n)}\,,
\quad
T_p=\sum_{n=-N_t}^{N_t} T_{p}^{(n)}\,,
\end{equation}
respectively; and the absorptance
as
\begin{equation}
A_p=1-(R_{p}+T_{p})\,.
\end{equation}

The rigorous coupled-wave approach~\cite{Faryad2,Li93,Moharam95,PMLbook} was used to compute the reflection and transmission amplitudes. As for the previous two sections, air was chosen as the
nondissipative partnering material and crystalline silicon as the dissipative
partnering material.  All calculations were made for $L_g=35$~nm, $L=350$~nm, $\zeta=0.5$, and $L_m=1000$~nm. We set $N_t=13$ after ascertaining that absorptance converged within a preset tolerance of $\pm1$\%.

\begin{figure}[!ht]
\begin{center}
\includegraphics[width=5.5in]{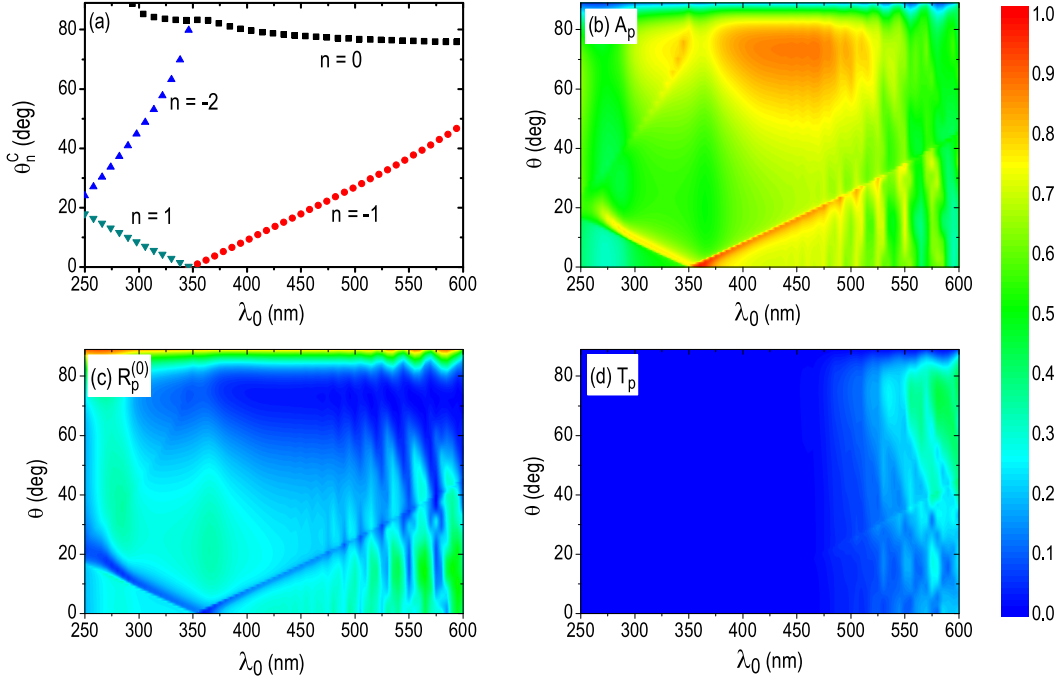}
\caption{
Practical configuration  with periodically corrugated guiding {interface.} (a)   Spectrums of $\theta_n^{\rm C}=\sin^{-1}\left\{\left[{\rm Re} (q)-2{\pi}n/L\right]/\ko{n_d} \right\}$ 
when the excitation of the Uller--Zenneck wave is
predicted by the solution of the canonical boundary-value problem.  (b) Absorptance $A_p$, 
(c) specular reflectance $R_{p}^{(0)}$, and (d) total transmittance $T_{p}$ as  functions of $\lambdao$ and $\theta$  when $L=350$~nm, $L_g=35$~nm, $\zeta=0.5$, and $L_m=1000$~nm.
}
\label{grating}
\end{center}
\end{figure}

According to Eq.~\r{dispsol}, an Uller--Zenneck wave could be excited when $\theta$ equals
\begin{equation}
\theta_n^{\rm C}=\sin^{-1}{\left\{\frac{ {\rm Re} (q)-2{\pi}n/L }{\ko{n_d}} \right\}}
\label{thetapeak}
\end{equation}
for some $n\in{\mathbb Z}$. Values of $\theta_n^{\rm C}$ for $n\in\left\{-2,-1,0,1\right\}$ are
 plotted in Fig.~\ref{grating}(a) as  functions of $\lambdao$.
 The absorptance $A_p$, the specular reflectance $R_{p}^{(0)}$, and total transmittance $T_{p}$ as  functions of $\lambdao$ and $\theta$ are presented in Figs.~\ref{grating}(b), (c), and (d), respectively.
The plot of $A_p$ in Fig.~\ref{grating}(b)  clearly shows the presence of sharp absorptance bands corresponding to $\theta^{\rm C}_n$  for $n=\pm1$  in Fig.~\ref{grating}(a) predicted by the canonical boundary-value problem. A somewhat less sharp absorptance band for $n=-2$ also exists. These absorptance bands  are independent of the
 thickness $L_m>2\lambdao$ and
indicate
the excitation of the Uller--Zenneck wave.

 A wide absorptance band in Fig.~\ref{grating}(b) delineated approximately by $\lambdao\in\left[375,525\right]$~nm and $\theta\in\left[50^\circ,80^\circ\right]$ does not represent the excitation of the Uller--Zenneck wave because, for any fixed value of $\lambdao$, the center of the
 $\theta$-range of this band is significantly different from $\theta_0^{\rm C}$ in
 Fig.~\ref{grating}(a). Moreover, the $\theta$-range is far too broad to signify the excitation of any surface wave.

Let us recall from Sec.~\ref{canonical} that the surface wave guided by the planar interface of air and silicon is an SPP wave for $\lambdao\leq293$~nm and an Uller--Zenneck wave for $\lambdao\geq294$~nm. A scan of Figs.~\ref{grating}(a), (b), and (c) suggests that the SPP wave blends into the Uller--Zenneck wave as $\lambdao$ increases, there being no discernible   difference between the pertinent absorptance bands across  $\lambdao\sim293.5$~nm.

The plot of the specular reflectance $R_{p}^{(0)}$ in Fig.~\ref{grating}(c) shows the signatures of the absorptance bands representing the excitation of the Uller--Zenneck wave for $n=\pm1$ in Fig.~\ref{grating}(b). Since the specular reflectance is not difficult to measure for $\theta\gtrsim10^\circ$, the grating-coupled configuration appears to be very appropriate for experimental confirmation of the existence of the
Uller--Zenneck wave. The selected combination of partnering materials---viz., air and silicon---is suitable
because Fig.~\ref{grating}(d) indicates that the total transmittance is very low in the chosen spectral regime due to the high absorption of light in silicon.

\section{Concluding remarks}\label{conc}
We investigated theoretically the excitation of the Uller--Zenneck wave guided by the interface of two homogeneous
and isotropic  dielectric materials, of which only one is dissipative. Although the solution of the
canonical boundary-value problem---involving the propagation of the Uller--Zenneck wave guided
by the interface of two contiguous half spaces  occupied by the two partnering materials---clearly
indicates the possibility of surface-wave propagation, a practical configuration involving the
planewave illumination of a planar
interface did not offer any corroborating evidence. It is therefore not surprising
that this configuration, although often used in the past
for experimental investigations of the Uller--Zenneck wave,   has not   yielded unambiguous confirmation of the existence of this type of surface wave.

In contrast, another practical configuration involving the planewave
illumination of a periodically corrugated interface of the
two partnering materials was shown by us to offer unambiguous confirmation of the
existence of the Uller--Zenneck wave. This surface wave can be excited as a Floquet
harmonic of order $n\ne 0$ for a wide range of the angle of incidence in the grating-coupled
configuration.

We hope that   this paper will set the stage for the first conclusive experimental evidence of the existence of the Uller--Zenneck wave. Such a development will be helpful in broadening the horizons of the applications of surface waves due to the availability of more choices of  partnering materials.

\vspace{0.5cm}

\noindent{\bf Acknowledgements.}  The authors thank the US National Science Foundation for partial financial support via grant DMR-1125590. AL also thanks the Charles Godfrey Binder Endowment at the Pennsylvania State University for ongoing support of his research activities.

%%%%%%%%%%%  Figure 1 begins %%%%%%%%%%
\newpage
%%%%%%%%%% Figure 1 ends %%%%%%%%%%%

%%%%%%%%%%%  Figure 2 begins %%%%%%%%%%
\newpage
%%%%%%%%%% Figure 2 ends %%%%%%%%%%%

%%%%%%%%%%%  Figure 3 begins %%%%%%%%%%
\newpage
%%%%%%%%%% Figure 3 ends %%%%%%%%%%%

%%%%%%%%%%%  Figure 4 begins %%%%%%%%%%
\newpage
%%%%%%%%%% Figure 4 ends %%%%%%%%%%%

%%%%%%%%%%%  Figure 5 begins %%%%%%%%%%
\newpage
%%%%%%%%%% Figure 5 ends %%%%%%%%%%%

 \end{document}